# Crowd tracking and monitoring middleware via Map-Reduce

Alexandros Gazis[a]* and Eleftheria Katsiri[a,b]

[a] *Department of Electrical and Computer Engineering, School of Engineering, Democritus University of Thrace, Xanthi, Greece ;*

*agazis@ee.duth.gr*

[b]*Institute for the Management of Information Systems, Athena Research & Innovation Center in Information Communication & Knowledge Technologies, Marousi, Greece*

This paper presents the design, implementation, and operation of a novel distributed fault-tolerant middleware. It uses interconnected WSNs that implement the Map-Reduce paradigm, consisting of several low-cost and low-power mini-computers (Raspberry Pi). Specifically, we explain the steps for the development of a novice, fault-tolerant Map-Reduce algorithm which achieves high system availability, focusing on network connectivity. Finally, we showcase the use of the proposed system based on simulated data for crowd monitoring in a real case scenario, i.e., a historical building in Greece (M. Hatzidakis' residence). The technical novelty of this article lies in presenting a viable low-cost and low-power solution for crowd sensing without using complex and resource-intensive AI structures or image/video recognition techniques.

**1 Introduction**

Map-Reduce is a data-parallel programming model used for processing and generating distributed computations of large data sets, as well as executing several clusters of data across commodity servers [1]. It provides a robust, scalable, and abstract method to process big data in parallel (i.e., large datasets) [2]. Specifically, this model provides a programming abstraction by expressing complex details of parallelization, fault-tolerance, data distribution, and load balancing in simple computations. Moreover, it provides data-intensive computing at a high level and acts as a cloud management tool for complex interconnected networking resources, such as storage and computers through the network's area [3].

In this article, we describe a middleware architecture for monitoring visitors in a preserved building using Raspberry Pi devices, i.e., small, affordable, customizable, and programmable mini-computers. Analytically, we have used an existing historical building as a case study, where we developed a system that studies the interconnectivity between several interconnected wireless sensor networks (WSNs) and mini computing devices. We clarify that we did not focus on spatial-location or energy management; instead, we studied the rationale of parallel computing to suggest a novice technique regarding a distributed fault-tolerant system, consisting of several low-power and low-cost mini-computers (e.g., Raspberry Pi).

It is important to note that while we chose to present a cloud-based system using WSNs, similar projects in indoor buildings typically adopt either AI solutions [4] or mobile phone sensing middleware [5]. Interestingly, in the last years, middleware has been used to monitor the structure monitoring of heritage buildings, including a medieval tower [6] and a historical building in Italy (Valentino Castle-Politecnico di Torino) [7]. Other notable projects which used indoor and outdoor middleware are: i) "CrowdWiFi" [8], locating open access Wifi points outside or inside buildings; ii) "iCrowd|" [9], focusing on continuous location updates for pilgrimage to Mecca; iii) "Urban Civics" [10], combining data on physical and social sensing from smart cities and urban scale models; iv) "SensorAct" [11], acting as a scalable middleware for privacy and security awareness on building management, v) "Cassowary" [12], studying smart content awareness in smart buildings and vi) "Openban" [13], providing data analytics for smart building with a high number of sensor data.

The technical novelty of this article is presenting a crowd monitoring application using a parallel algorithm executed by mini computers. Additionally, in the context of a pandemic, these results may support the tracking of crowd density in buildings to minimize viruses' spread. The output results of this operation can either be used as an evacuation technique, according to [14] or as a tool to pinpoint which rooms and exhibitions attract the most visitors, similarly to the techniques described by [15]. We highlight that unlike similar solutions proposed in current bibliography, our middleware is less resource-intensive because it does not perform data-intensive tasks (e.g., neural network or image/video analysis). We believe that the most novel aspect of this article does not lie in providing a detailed bottom-up technical approach on problems, such as sensors' calibration, network protocols, etc. but rather in presenting a top-down middleware solution.

In this paper, we first define and explain the terms "middleware", "Map-Reduce" and our case study (i.e. a historical building in Greece). Then, we focus on presenting our rationale regarding the experimental setup. Specifically, based on the architectural layer of the proposed middleware (see Figure 1), we discuss common issues regarding server-client remote connection the used hardware, sensors, and RFID tag frequencies. Moreover, we describe the Map-Reduce framework used to develop our remote system. Lastly, we provide benchmarks of the proposed system and algorithm, and we discuss our findings and conclusions.

**2 Background**

In this section, we briefly present the key concepts of this publication. Specifically, we clarify the meaning of the term "middleware" and we present our case study for developing an IoT middleware application. Then, we introduce the Map-Reduce application used.

*2.1 Middleware*

The term "middleware" was first used in the early 1980s to describe the software solutions proposed for upgrading legacy systems—mainframes—and migrating them to more advanced technologies and software frameworks. As stated in [16], we characterize middleware as "those services found above the transport (i.e., over TCP/IP) layer set of services but below the application environment (i.e., below application-level APIs)." Middleware typically exists between software applications and the operating system as an intermediary layer. It is responsible for acting as a message broker (e.g., point to point, publish-subscribe) and is largely used in structuring the logic behind back-end development software applications. For example, the simplest middleware developed is when a server sends a request, and the client gets that request and produces a response (action).

The middleware architecture developed for crowd monitoring in a preserved building consists of four layers (see Figure 1). Specifically, these are the following:

1. A physical layer consisting of mini-computer devices, such as a Raspberry Pi or other (mini) computer devices.

2. A cloud layer consisting of basic communication functionalities, such as the database connection and interaction (query execution) to detect the server-client IP address and exchange information via messages, using the User Datagram Protocol (UDP).

3. An application layer (software and communication services) implementing Map-Reduce algorithm and data analysis for rooms and visitors.

4. A data layer extracting, transforming, and loading data measurements into (text) files and local variables.

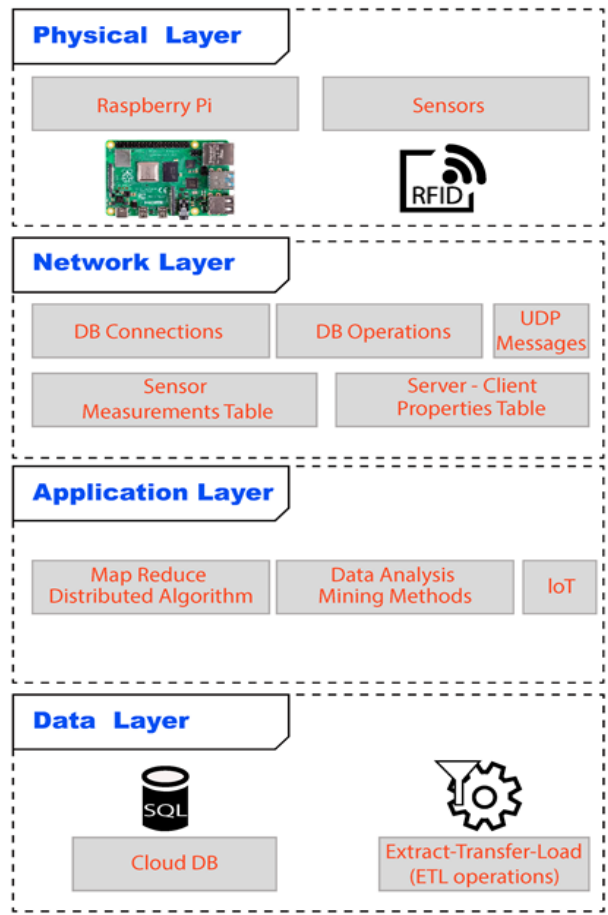

**Figure 1**. The architectural layer of our crowd monitoring middleware.

## 2.2 Case Study: Hatzidakis' House

Our model of study was a building of high significance in the city of Xanthi in Greece. The house (presented in Figure 2) was the residence of M. Hatzidakis, one of the greatest Greek music composers and it was constructed in 1829. Additionally, the house which currently operates as a cultural center, has a Baroque-style with neo-classical elements, consisting of three floors and covering a plot of 1.317 sq. m. [17, 18]. The floor plan is presented in Figure 3 and radio frequency identification (RFID) readers are shown in red. We note that besides the tests and scope of this publication, the only room that will not be monitored will be the floor's WC to ensure people's privacy.

Specifically, Hatzidakis' house was used to validate the system via simulated data from a distributed visitor counting. We simulated our testing on the first floor which hosts art events and exhibitions. The integrated system was designed as an IoT use-case to provide an effective and reliable method to count, track, and monitor visitors.

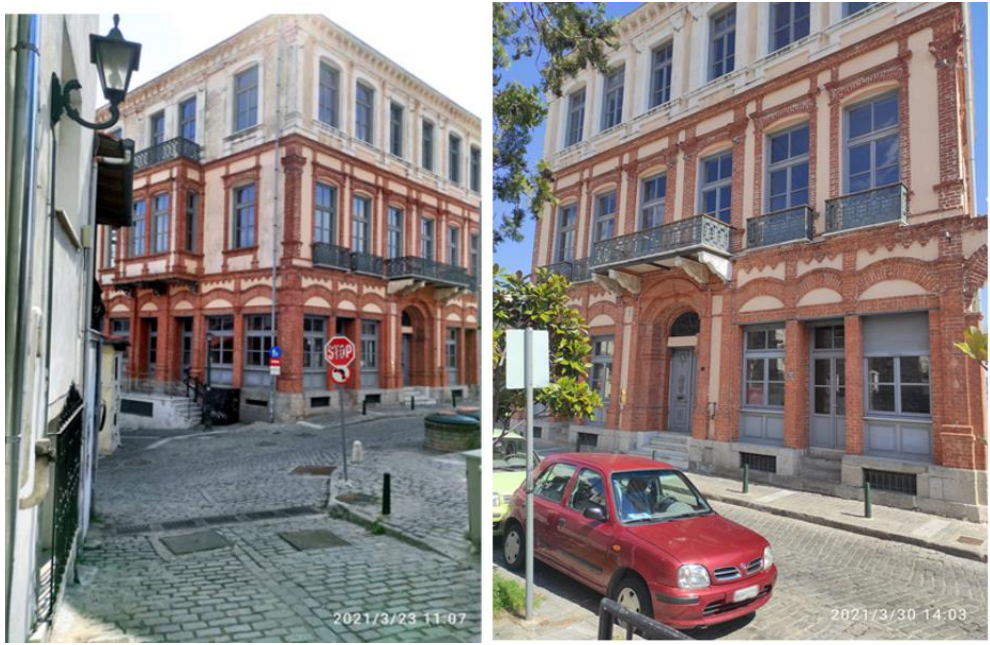

**Figure 2**. Outside view of M. Hatzidakis' residence.

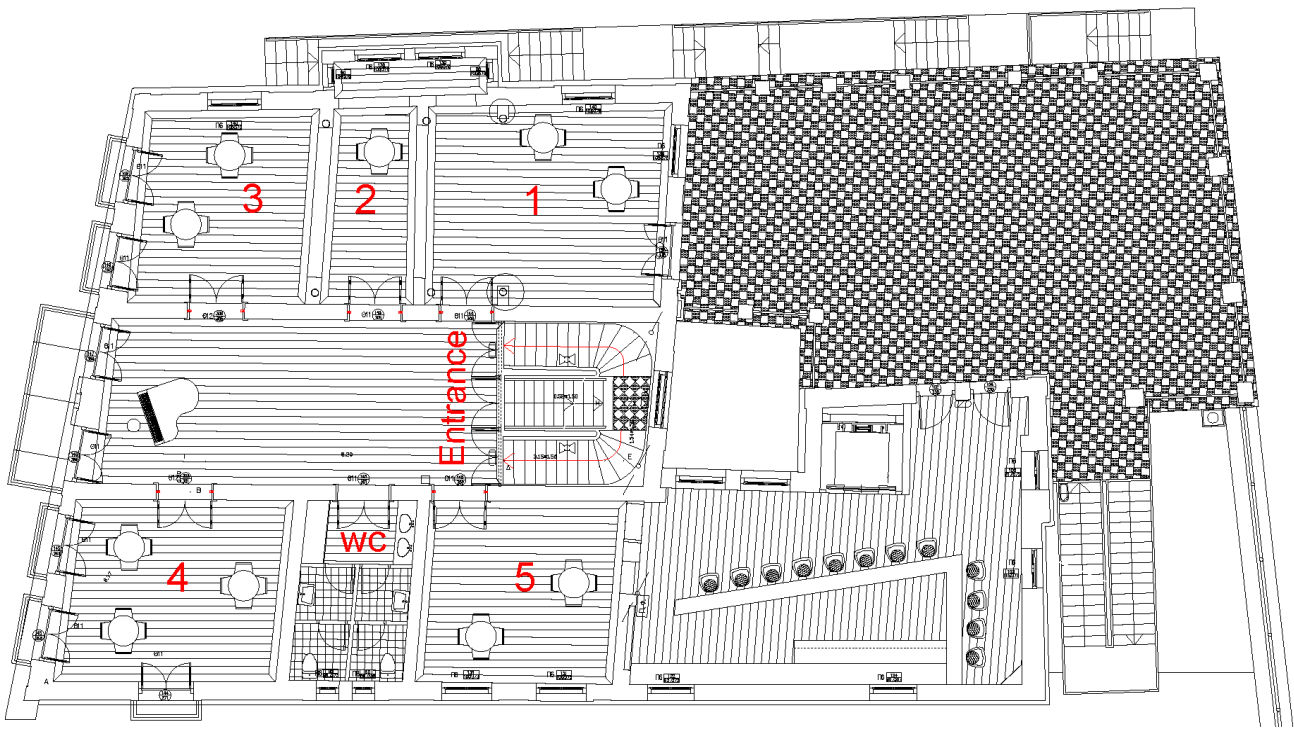

**Figure 3**. Floor plan of Hatzidakis' residence.

## 2.3 Experimental Setup

In this section, we provide information regarding the experimental setup of our proposed middleware. Analytically, visitors entering the building are provided 1 of the 3 available RFID tags, encoding information on whether they are men, women or animals. We clarify that producing an animal tag was necessary for tracking animals used by persons with disabilities. Moreover, each room is equipped with a small IoT device (room computer) that is connected to a RFID reader. Room computers (also referred to as "workers") are threaded and they communicate with the server using sockets. Workers may implement two roles; they are either mappers or reducers. On the one hand, mappers are responsible for scanning the room periodically to count the partial number of visitors that are in the room. On the other hand, reducers count the data that has been assigned to them by the Master, summing up all occurrences of passers and returning the data to the server.

Furthermore, we used Raspberry Pi model 4, i.e. a device of 4GB of RAM equipped with an ARM Cortex BCM2711 at 1.5 GHz. We note that similar results can be achieved via model 3B or marginally 3A+ respectively, as mentioned in the following sections. Moreover, we suggest the placement of 2 RFID readers on the entrance and each door (one on the left and one on the right of each gateway) to identify passers. As for the RFID tags, we recommend the use of waterproof and dust-resilient wristbands. To operate frequencies up to 134 kHz, we may achieve a view of sight of approximately 20-30 cm. As for higher frequencies up to 13.56 MHz, the view of sight is around 1 meter.

Lastly, we took into consideration that the average person's width is 35-41 cm, according to the Center for Disease Control and Prevention (CDC) [19], while entrances and doors of the studied building have an average width of 1 meter [20]. Thus, we calculated the hand movement of passers when walking and recommend the use of tag frequencies close to 134 kHz. We note that, although higher frequencies could provide better monitoring accuracy, we believe that the trade-off between cost and accuracy is not extreme.

## 2.4 Description of Map-Reduce use

Map-Reduce is an extensively researched model [21], mainly due to widely used open-source projects, such as Spark and most notably Hadoop [22]. The Hadoop architecture, defined as a master-worker architecture, consists of a master node (Name Node) that manages, maintains, and monitors a large number of workers, i.e., worker nodes

(Data Nodes). Map-Reduce is the core component of Hadoop that processes a considerable amount of data in parallel by dividing the work into a set of independent tasks and by being responsible for their tracking (Task Tracker) and overall coordination (Job Tracker). Job Tracker and Name Node define the master's behaviour while Data Node and Task Tracker define the worker's behaviour [23].

Based on the above, the "master" is a single point of failure in the Hadoop architecture. The original Map-Reduce framework aborts the Map-Reduce computation when the master fails, while Hadoop provides a solution based on check pointing. As a result, when the master fails, it can be respawned from the latest checkpoint. The problem with this approach is that "check pointing" it is not a feasible option for a sensor network. Specifically, in this type of networks, nodes usually have limited persistence to store large checkpoint information (apart from SD cards, which are limited in size).

Additionally, even if we increase the storage capability to store more information, these networks only support a limited number of rewrites, and in this case, the nodes themselves are at risk of failing altogether. This results in another (available) node of the system taking over and coordinating the nodes.

We have chosen parallel processing to address issues such as limited storage space capabilities and hardware errors regarding data rewrites. Moreover, distributed processing is less resource-intensive and with lower overall cost as it can be implemented with low-power and low-cost components.

In respect to the application layer, we suggest a novice application of a Map-Reduce algorithm where a "master-worker" IoT architecture is described that guarantees liveness, safety, and high availability. This application focuses on monitoring the visitors in a historical building (see Figure 4). When entering the preserved building, the visitor will be provided a RFID tag based on their sex or species (pets will be provided with a specific tag).

Several low-power and low-cost computer devices that gather this RFID information will be placed in each floor, typically one per room. Each device communicates remotely with a database and executes the Map-Reduce algorithm to extract information based on the RFID tags.

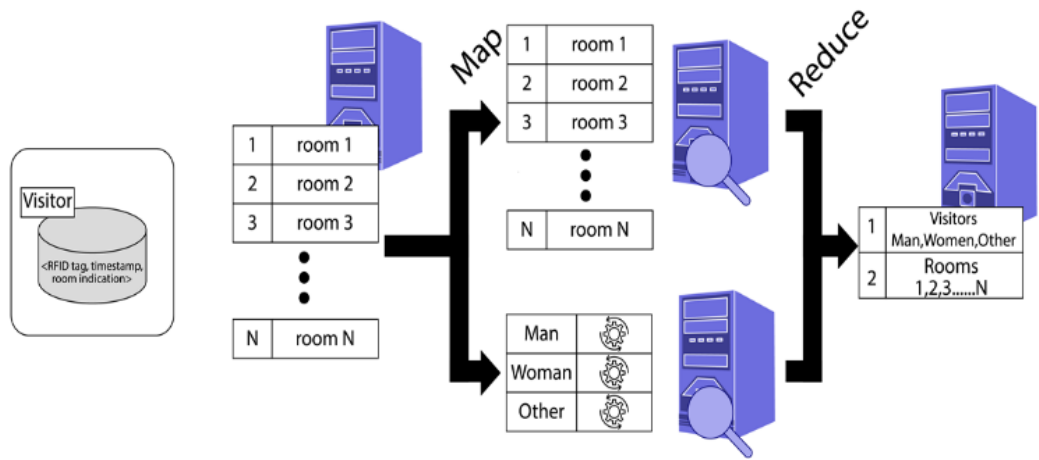

**Figure 4**. Map Reduce algorithm proposed for crowd monitoring.

Except for the Map-Reduce visitor counter application, we also developed an algorithm allowing nodes to act either as clients or servers in order to overcome known issues, such as Hadoop's master single point of failure. The criterion for selecting a leader was its degree of connectivity to the network, i.e., that it communicates with at least two-thirds of the total nodes. In case more than one candidate leader satisfied this constraint, then the current leader-server retains its status.

**4 Algorithm Proposed**

To create a solution that would be more diverse and would incorporate "real" computer devices, we developed our application as follows:

1. Server-Client properties must only be provided during the first execution. Specifically:

    a. The user states the exact number of clients that will be connected.

    b. The system's Server-Client's Socket unique information (IP/Port) should be

known at all times.

   c. The specific computer device that acts initially as a server must be pinpointed.

2. Server-Client nodes must be capable of exchanging roles to eliminate cases like:

   a. The server losing the connection (e.g., due to poor network).

   b. The server being temporarily unavailable (e.g., due to power problems, damaged equipment, etc.).

3. Every computer device connected has a unique node number to easily address them individually.

4. In a given timeframe (e.g., 1 minute), Step 2 and 3 should occur, i.e., the system's connectivity and the server's operational fitness should be checked.

5. In a given timeframe (e.g., after a guided tour of 5 to 15 minutes), all workers and the master must be able to perform a set of tasks, such as uploading given room measurements to the remote database (DB), checking connectivity, crowd-monitoring (visitors) via Map-Reduce algorithm, etc.

6. The application should be executed endlessly until all computer devices fail to operate.

Our algorithm focuses on role exchanging as part of a system's built-in functionality. A flow diagram of the six steps discussed is presented in Figure 5.

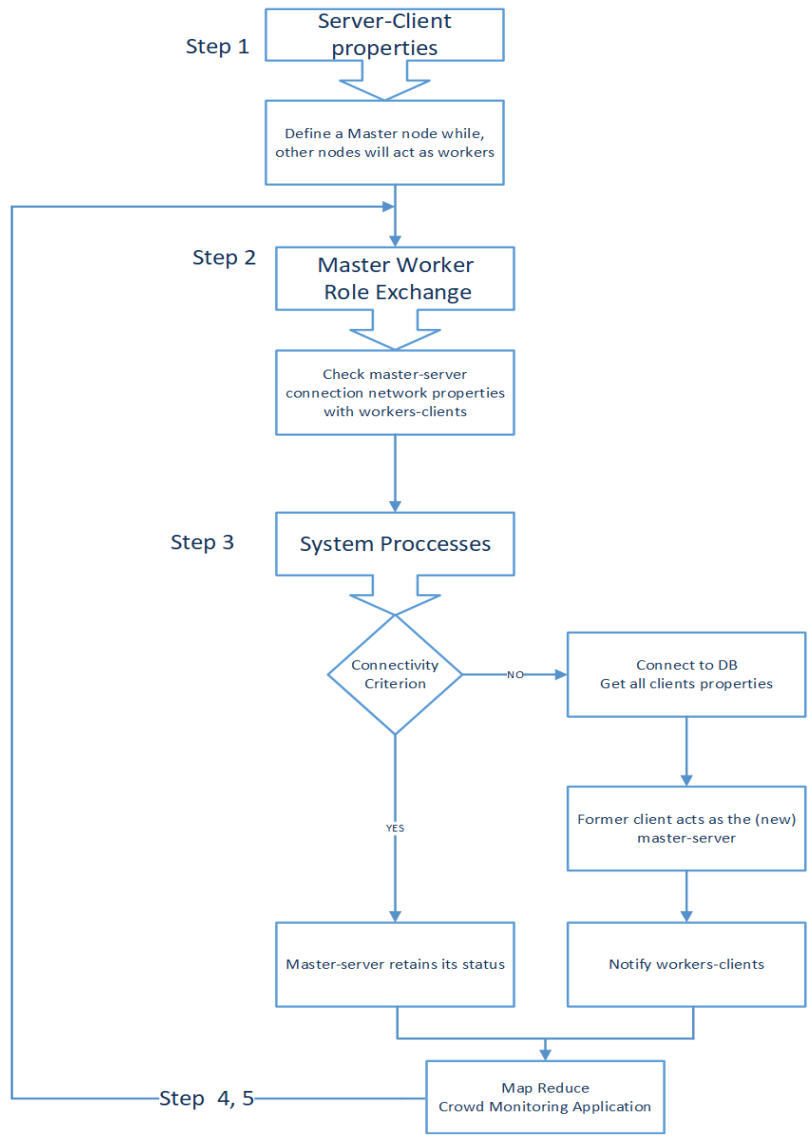

**Figure 5.** Flaw diagram of our proposed middleware's operating principles.

## 5 Results

For each input data stream, whether that is a text file or data from our database, the system executes the Map-Reduce algorithm as follows:

- Case 1: < ($key_1$, $value_1$) = (visitor, value) >

- Case 2: < ($key_2$, $value_2$) = (room, value) >

For example, regarding the Hatzidakis' residence, we selected the ground floor as our model of study, i.e., a quad-core mini-computer of 4 GB RAM-size (Raspberry Pi model 4B) was placed in each room, thus simulating a total of 5 computer devices' data. For an input data stream from a guided tour of 500 visitors which timespan was approximately 15 minutes, the output results of our system (as presented in the terminal) are shown in Tables 1 and 2 for each room and visitor monitoring, respectively. Additionally, in Figure 6, we provide the execution times of the Map-Reduce algorithm for a broad number of visitors.

Moreover, in Table 3, we provide measurements regarding the **application layer** of our application. Specifically, we have measured the CPU usage, memory usage, and execution time regarding the boot up (initiation) of the server-client-side communication, i.e., the CPU usage, memory usage and power consumption. These measurements occur for the period of the server and client to boot, establish a socket connection, and exchange the first successful UDP messages. Additionally, in Table 4, we provide the execution time end-to-end for a full software cycle (initial execution of our system up until it completes step 1 to -5 of our proposed algorithm). It is noted that measuring the performance of each class was not easy. There are several faults in the measurements, including the JVM platform internal parameters, like the JIT's compiler optimizations in the outcome.

Lastly, as for the **networking layer** of our application, we provide measurements regarding the average response times for the server and client modules. Our system is capable of sending and receiving UDP messages between the server and the client. Currently, these messages provide information regarding the socket connection and connectivity status, meaning they do not exchange the system's information (e.g., sensor measurements). Specifically, we provide the average time for our cloud tier, i.e., for a request to be sent (client) and be successfully accepted by a recipient (server). The time responsiveness of our system is presented in Figure 7.

**Table 1.** Map-Reduce algorithm for case 1: visitor monitoring (man, woman, other).

```
>>Sum: { Man : 157, Woman : 169, Other : 174 }

{ Other-room5=28,   Other-room4=37,   Other-room3=31,
  Other-room2=16,   Other-room1=31,   Other-room0=31,
  Woman-room0=28, Woman-room1=31, Woman-room2=36,
  Woman-room3=23, Woman-room4=23, Woman-room5=28,
  Man-room1=22,    Man-room0=27,    Man-room3=30,
  Man-room2=28,    Man-room5=28,    Man-room4=22 }
```

**Table 2.** Map-Reduce algorithm for case 2: room monitoring (1,2,3,4,5).

```
>>Sum:{ Room0 : 86, Room1: 84, Room2: 80, Room3 : 84, Room4: 92, Room5:84 }

0: { Man: 27, Woman: 28, Other: 31 }   1: { Man: 22, Woman: 31, Other: 31 }
2: { Man: 28, Woman: 36, Other: 16 }   3: { Man: 30, Woman: 23, Other: 31 }
4: { Man: 22, Woman: 23, Other: 37 }   5: { Man: 28, Woman: 28, Other: 28 }
```

**Table 3.** Server-Client communication measurements.

|  |  | Server | Client |
|---|---|---|---|
| CPU | [%] | 59 | 40 |
| Memory | [MB] | 205,7 | 196,9 |
| Power Consumption | [A] | 0,52 | |
| | [V] | 5,31 | |

**Table 4.** Server-Client measurements for an entire software cycle.

| Visitors |  | 50 | 250 | 500 | 1000 |
|---|---|---|---|---|---|
| CPU | [%] | 41,20 | 52,43 | 60,41 | 67,60 |
| Memory | [MB] | 845,88 | 866,60 | 892,24 | 990,89 |
| Power Consumption | [A] | 0,80 | 0,81 | 0,83 | 0,84 |

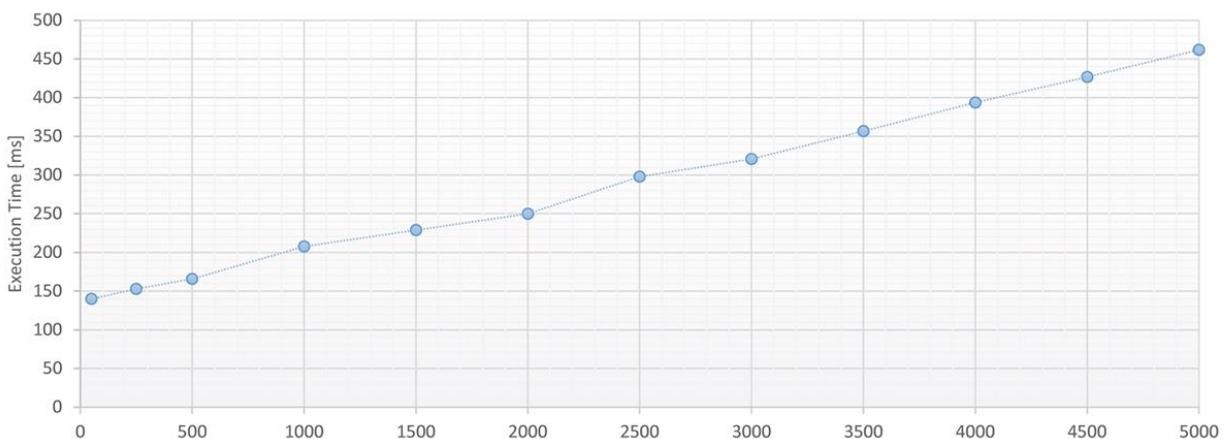

**Figure 6.** Map-Reduce algorithm for case 2: room monitoring (room 1 to N).

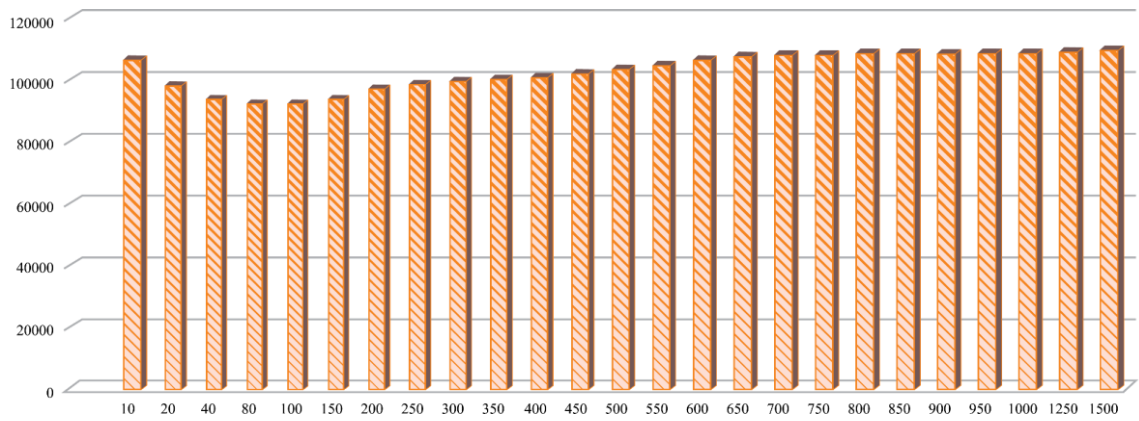

**Figure 7.** Average time (in msecs) for server-client response for 0 to 1500 requests

## 6 Conclusions

In this paper, we presented a middleware that can use IoT receiving data derived from a WSN. Our system task is to perform crowd tracking and monitoring. Moreover, the output regarding building visitors can be used to assist evacuations, in case of an emergency or as a tool to pinpoint which rooms and exhibitions hold the greatest interest or to track people's numbers in the building to minimize a virus spread (e.g., COVID-19).

All layers of our system use open-source solutions running on mini low-power and low-cost computers, such as Raspberry Pi. Specifically, the largest part of our application is cloud-based, reliable, and not resource intensive. In our experiments, we mostly used new versions of Raspberry Pi (4), but similar results can be achieved via older versions such as 3, which has 1GB of RAM, or even systems with lower memory capabilities.

The novelty of this paper lies in presenting the algorithmic approach (i.e. the steps) of the application layer. As evident, multiple technologies were utilised to develop this middleware application, including Map-reduce, WSN and crowd sensing. The suggested IoT middleware combines many different aspects of computing, focusing on simplifying the processes and the overall computing cost of a crowd sensing middleware, because the majority of the state-of-the-art literature relies either on AI or image analysis. Although these approaches provide similar (or better results), we believe that the trade-off between crowd monitoring accuracy and devices' cost is notable.

Similarly to most applications executing Map-Reduce algorithms, our proposed application has high availability and virtually little downtime. While the endless model proposed is not resource-effective, it does not cause system failures. the proposed system—consisting of many minicomputers—is an interesting case study, as it does not use high-end hardware for the server. On the contrary, it proposes a fairness index in the application's processing. Lastly, our approach can be used either as a standalone solution or as a means to cheaply and reliably validate the crowd monitoring results of more expensive applications.

In future steps of this work, we need to focus on implementing a leader election algorithm. Specifically, a distributed algorithm must be developed to help select a process among a set of processes that will act as the leader. This leader will be responsible for the coordination of the next steps among the processes. Leader election is particularly important in distributed systems, such as sensor networks where there is no central control and each node does not know their role in the network. This way, a fair allocation of roles and tasks can be achieved.

**Acknowledgements**

This article details our recent work, as part of an MSc course on distributed systems, at the School of Engineering Department of Democritus University of Thrace, Greece to apply the parallel programming to WSNs. Lastly, the authors are grateful to the civil engineer E. Chamalidou for her assistance in the analysis and the presentation of the floor plans of M. Hatzidakis' residence, as presented in this publication.